\documentclass[aps,pra,reprint,superscriptaddress]{revtex4-2}
\usepackage{amsmath} 
\usepackage{amssymb} 
\usepackage{physics} 
\usepackage{graphicx} 
\usepackage{hyperref} 
\usepackage[version=4]{mhchem} 
\usepackage{bm} 
\usepackage{color,soul} 

\makeatletter
\newcommand{\colorcaption}[2][]{%
  \begingroup%
  \renewcommand{\@caption@fignum@sep}{ (Color online). }%
  \caption[#1]{#2}%
  \endgroup%
}
\makeatother

\begin{document}

\title{Electronic control and switching of entangled spin state using anisotropy and exchange in the three-particle paradigm}

\author{Eric D. Switzer}
\affiliation{Department of Physics, University of Central Florida, Orlando, Florida 32816, USA}
\author{Xiao-Guang Zhang}
\affiliation{Department of Physics, Center for Molecular Magnetic Quantum Materials and Quantum Theory Project, University of Florida, Gainesville, Florida 32611, USA}
\author{Talat S. Rahman}
\email[Corresponding author email: ]{talat.rahman@ucf.edu}
\affiliation{Department of Physics, University of Central Florida, Orlando, Florida 32816, USA}

\date{\today}

\begin{abstract}
We explore the control and switching of the entangled spin states of multi-spin particle qubit coupled to an electron using a three-particle spin model described by $S_i$ ($i=1,2,3$), in which $S_1=\tfrac{1}{2}$ is an electron and $S_{2,3}$ can have any spin with both exchange coupling and magnetic anisotropy. We derive a general formula for the existence of a switching (DJ) resonance for \textit{any} spin $S_{2,3}$. We further contrast the entanglement switching mechanisms for the $S_{2,3}=\tfrac{1}{2}$ and $S_{2,3}=1$ spin models. We find that while the onsite magnetic anisotropy in the case of $S_{2,3}>\tfrac{1}{2}$ allows full control of their spin states via interaction with $S_1$, in order to achieve acceptable control of a Bloch vector within the $S_{2,3}=\tfrac{1}{2}$ model, additional mechanisms, such as anisotropic exchange coupling, are required.
\end{abstract}

\maketitle

\section{Introduction}
    The ability to generate and stabilize entanglement within qubits is a prerequisite for usable quantum information devices.
    Spin qubits, which are the coherent superposition of spin states within quantum objects, make use of entangled spin states for quantum gate operations \cite{nielsenbook}. 
    This general class of qubits has had much success and is seen as a promising candidate for scalable quantum information science (QIS) technologies \cite{chatterjee2021}.
    Some applications, like those utilizing confined electrons in quantum dots fabricated in semi-conductors, use manipulation of electrostatic gates, electric dipole spin resonance, and applied magnetic fields to generate, stabilize, and manipulate the entangled states \cite{loss98,petta05,hanson07,noiri16,noiri18,nakajima19,yang2020,leon2020}.
    
    Some of these preparation methods have also been applied to QIS approaches that utilize molecular magnets \cite{leuenberger01, vandersypen01,vincent12,ganzhorn13,urdampilleta13,thiele14,pedersen16,najafi19}. 
    Molecular magnets, such as $\text{TbPc}_{2}$ \cite{urdampilleta13} and $\text{Mn}_{12}$ \cite{gatteschi2003}, are complex molecules that possess an onsite magnetic anisotropy because of their larger magnetic moments that distinguishes them from the $S=\tfrac{1}{2}$ Ising spins. 
    The long coherence times, the ability to tunnel between spin states enabled by their magnetic anisotropy, and tailorable ligands \cite{bogani08} make molecular magnets desirable candidates for QIS systems.
    
    In several QIS approaches, exchange coupling plays a key role \cite{cronenwett98,leuenberger06,gonzalez08}, including the use of the Kondo effect \cite{kondo64} in the switching of entanglement states \cite{tooski2014}. 
    Considering magnetic anisotropy in molecular molecules, it is important to incorporate both exchange coupling and magnetic anisotropy within models to predict and realize spin entanglement scenarios.
    For example, the study of an electron scattering off of two magnetic impurities under a contact exchange interaction found coherent transmission scenarios based on the entangled state of the impurities \cite{costa06,ciccarello06,ciccarello09}.
    The study of the interplay of exchange coupling and magnetic anisotropy goes beyond QIS, and has applications in systems which contain both of these dynamical interactions as in ultra-cold optical lattices \cite{chung2021}. 
    While the study of these two effects has been explored for two particles \cite{zitko08}, the community's drive to realize scalable QIS devices may require the investigation of multi-particle models similar to those explored in Ref.~\cite{mehl15}.
    Work that incorporates multiple sites/particles, exchange coupling, and magnetic anisotropy is ongoing \cite{hiraoka17,switzer2021}.
    
    In this work, we further investigate a general model introduced in Ref.~\cite{switzer2021} that incorporates the interplay of exchange coupling with magnetic anisotropy for three stationary spin particles. 
    In the interest of maintaining wide applicability to contexts outside of QIS, we consider a general model in which two of the spin particles can possess any spin magnitude.
    Similar to the concepts used in quantum dot technologies and ultra-cold optical lattices, we neglect the kinetic components of the physical system, and concentrate solely on the spin degree of freedom.
    
    As we will show, we find the interactions of the exchange coupling and magnetic anisotropy terms leads to a set of conditions that correspond with perfect (lossless) switching between non-entangled and entangled states, named as ``DJ resonances,'' for any spin $S_{2,3}>\tfrac{1}{2}$.
    These resonances allow for full control of appropriately chosen Bloch vectors. 
    We contrast these resonances for two applications of the general model: one in which all three particles possess a spin $S_{1,2,3}=\tfrac{1}{2}$, and another in which two of the particles possess a spin $S_{2,3}=1$.  
    We also demonstrate that control of the entanglement state can be accomplished by appropriate spin filtering of the $S_{1}$ particle for any spin $S_{2,3}$.

\section{Theoretical Method}
\begin{figure}
    \includegraphics[width=\columnwidth]{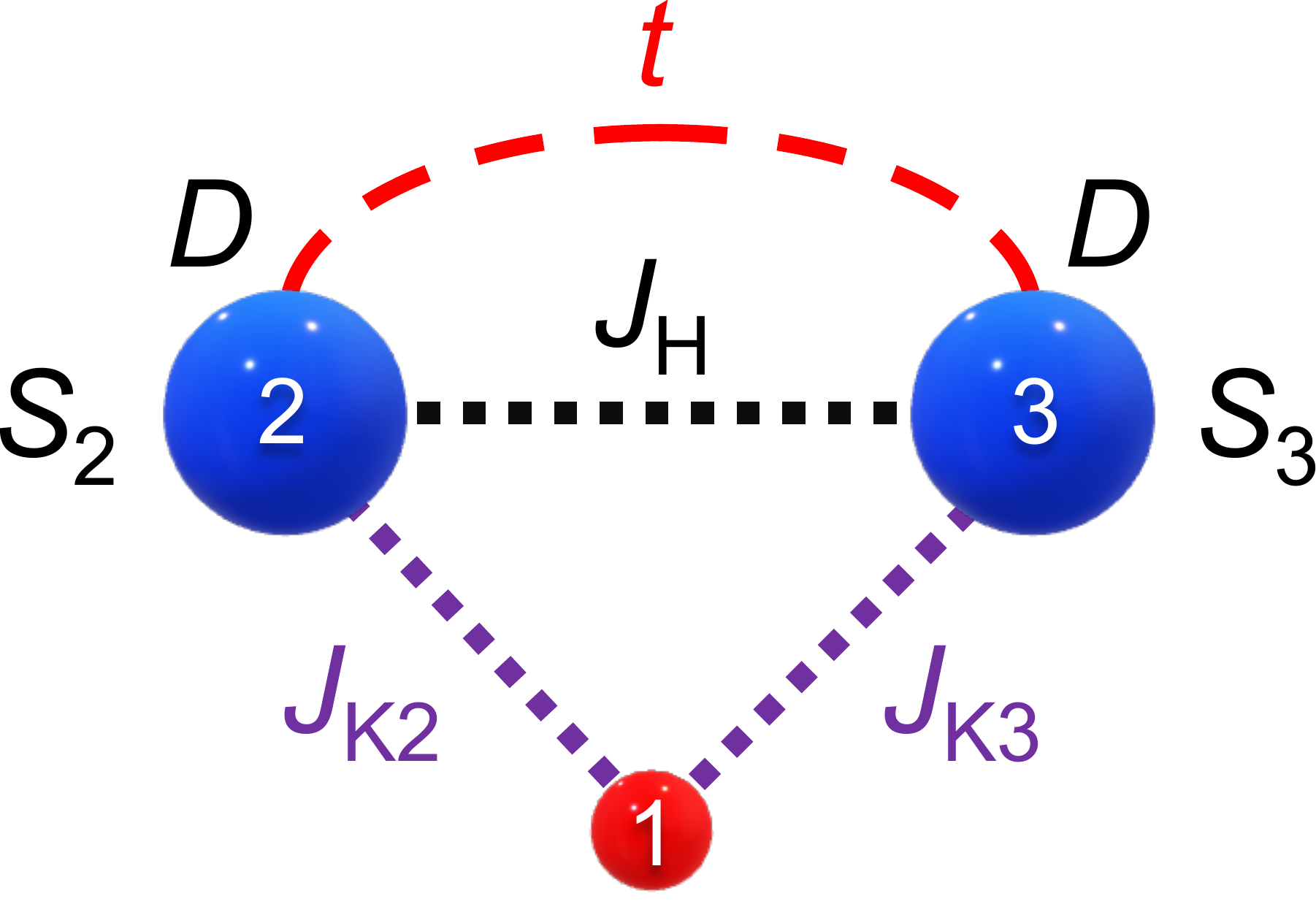}
    \colorcaption{\label{fig:schematic}
    Schematic of the three particle spin model. 
    Particle 2 and 3 are coupled via an exchange interaction $J_{\text{H}}$, and possess magnetic anisotropy $D$. 
    Each couple with particle 1 with an exchange interaction $J_{\text{K2}}$ and $J_{\text{K3}}$, respectively. 
    Particle 1 may hop between particle 2 and 3 with hopping strength $t$.
    }
    \end{figure}

    As shown in the schematic in Fig.~\ref{fig:schematic}, we consider four interactions: the similar uniaxial magnetic anisotropy possessed by particle 2 and 3, the Heisenberg-like exchange interaction between particle 2 and 3, a Kondo-like interaction between particle 1 and 2 and particle 1 and 3, and a hopping term which describes the movement of particle 1 between particle 2 and 3.
    We also allow the parameters of the system to generically encompass all available regimes. 
    This is achieved by both allowing the exchange interaction and magnetic anisotropy parameters to take on all values. 
    In other words, the exchange interactions are allowed to be ferromagnetic ($J<0$) or antiferromagnetic ($J>0$), and both hard ($D>0$) or easy ($D<0$) magnetic anisotropy axes are permitted.
      
    \subsection{Model Hamiltonian}
    
        The total Hamiltonian is,
        \begin{align}
        \label{eqn:hamiltonian-total}
            \mathcal{H} &= \mathcal{H}_{\text{H}}
                          +\mathcal{H}_{\text{K}}
                          +\mathcal{H}_{\text{A}}
                          +\mathcal{H}_{\text{t}},
        \end{align}
        where $\mathcal{H}_{\text{H}}$ is the Heisenberg-like Hamiltonian, $\mathcal{H}_{\text{K}}$ is the Kondo-like Hamiltonian, $\mathcal{H}_{\text{A}}$ is the magnetic anisotropy Hamiltonian, and $\mathcal{H}_{\text{T}}$ is the hopping Hamiltonian.
        From this point forward, $\hbar$ = 1. 
        The Heisenberg-like interaction takes the form,
        \begin{align}
        \label{eqn:hamiltonian-heisenberg-interaction}
            \mathcal{H}_{\text{H}} = 
            J_{z}\hat{S}^{z}_{2}\hat{S}^{z}_{3}+J_{xy}\left(\hat{S}^{x}_{2}\hat{S}^{x}_{3}+\hat{S}^{y}_{2}\hat{S}^{y}_{3}\right),
        \end{align}
        where $\hat{\vb{S}}_{i}=(\hat{S}^{x}_{i},\hat{S}^{y}_{i},\hat{S}^{z}_{i})$ is the spin operator for the $i$th particle, $J_{z}$ is the strength of the exchange interaction between particle 2 and 3 parallel to the direction of the magnetic anisotropy axis, and $J_{xy}$ is the strength of the exchange interaction between particle 2 and 3 perpendicular to the direction of the magnetic anisotropy axis. This exchange term is similar to those used to describe the exchange interaction between two coupled dimers with an Heisenberg XXY model \cite{hill2003}. When this interaction is taken to be isotropic, i.e. $J_{z}=J_{xy}\equiv J_{\text{H}}$, this equation simplifies to $\mathcal{H}_{\text{H}}=J_{\text{H}}\hat{\vb{S}}_{2}\cdot\hat{\vb{S}}_{3}$.
        The exchange interaction of the $S_{1}$ particle with the $S_{2,3}$ particles is closely related to the spin portion of the Kondo interaction, and may be represented by,
        \begin{align}
        \label{eqn:hamiltonian-kondo-interaction}
            \mathcal{H}_{\text{K}}=\frac{1}{2}\sum_{\mu,\mu^{\prime},i=2,3}J_{\text{Ki}}\hat{\vb{S}}_{i}\cdot\hat{d}^{\dagger}_{\mu,i}\hat{\bm{\sigma}}_{\mu,\mu^{\prime}}\hat{d}_{\mu^{\prime},i},
        \end{align}
        where $\mu$ is a spin index for particle 1, $\bm{\sigma}_{\mu,\mu^{\prime}}$ is the corresponding $\mu,\mu^{\prime}$ matrix element of the $s=\frac{1}{2}$ Pauli matrix, and $\hat{d}^{\dagger}_{\mu,k}$/$\hat{d}_{\mu,i}$ represents (in second quantization language) the creation/annihilation operator of a state in which particle 1 is bound to particle $i$. 
        
        The magnetic anisotropy term is given as,
        \begin{align}
        \label{eqn:hamiltonian-anisotropy}
            \mathcal{H}_{\text{A}}=D\left(\hat{S}^{z}_{2}\hat{S}^{z}_{2}+\hat{S}^{z}_{3}\hat{S}^{z}_{3}\right), 
        \end{align}
        where $D$ is a uniaxial anisotropy strength. 
        This Hamiltonian term is derived differently in depending on the context of the spin particle type. For example, if the two spin particles refer to magnetic molecules, magnetic anisotropy is explained by geometric distortions of constituent ions \cite{gatteschi2003}. Conversely, in  ultra-cold optical lattices, an effective magnetic anisotropy is created by direct on-site interactions between atoms in two states \cite{chung2021}. We note that the magnetic anisotropy Hamiltonian term is only meaningful for $S_{2,3}> \tfrac{1}{2}$.
        
        Last, the hopping Hamiltonian is described by,
        \begin{align}
        \label{eqn:hamiltonian-hopping}
            \mathcal{H}_{\text{t}} = 
            \sum_{\mu}\left\{t\hat{d}^{\dagger}_{\mu,2}\hat{d}_{\mu,3}+h.c.\right\},
        \end{align}
        where $\mu$ is the spin index for particle 1. This term describes kinetic motion of the $S_{1}$ particle between $S_{2}$ and $S_{3}$ and vice versa. As will be shown later, the hopping term does not play a significant role in the spin dynamics of the system. This, however, does not mean that the hopping term serves no role. We show later that the spin dynamics form a necessary condition in measuring the entanglement state of the coupled particles, but this condition is not a sufficient one. To realize the results of this work in an experimental setup, one should also include the spatial degrees of freedom of the spin particles. For example, we give a derivation of this term as an extension of the two-site Anderson impurity model in the supplementary material, in which the momentum components of this term will certainly contribute to the overall dynamics of that system. While not vital for those systems that do not fit the impurity model, we conclude that it is important to keep the hopping term due to the possible role that the $S_{1}$ particle will play as a transient carrier of information of the coupled particles' entanglement state. Because of the diagonal nature of the term when projected onto spin space, one may remove the term if it is not needed. 
        
    \subsection{Dynamics and Choice of Basis}    
        To examine the dynamics of the system, we use the language of density operators to analytically and numerically solve the Liouville-von Neumann equation (the density operator equivalent of the time-dependent Schr\"{o}dinger equation),
        \begin{align}
        \label{eqn:liouville-von-neumann}
            i\pdv{\rho}{t}=\left[\mathcal{H},\rho\right],
        \end{align}
        where $\rho$ is the density operator in the Schr\"{o}dinger picture, and the brackets denote the commutator. 
        In general, the solution for Eq.~(\ref{eqn:liouville-von-neumann}) is,
        \begin{align}
        \label{eqn:time-dependent-density-matrix}
            \rho(t)=U(t)\rho(0)U^{\dagger}(t),
        \end{align}
        where $U(t)$ is the unitary time evolution operator,
        \begin{align}
        \label{eqn:unitary-time-evolution-operator}
            U(t) = \mathcal{T}\left[\exp\left(-i\int_{0}^{t}\mathcal{H}(\tau)\dd{\tau}\right)\right],
        \end{align}
        and $\mathcal{T}$ is the time-ordered operator. 
        For the familiar case in which the Hamiltonian is time-independent, like in all of our applications to our model considered in this paper, the unitary time evolution operator simplifies to,
        \begin{align}
        \label{eqn:unitary-time-evolution-operator-hamiltonian-time-independent}
            U(t) = e^{-i\mathcal{H}t}.
        \end{align}
        Thus, in principle, the time-dependent behavior of our model's density matrix can be solved exactly.
        
        It is important to consider specific choices of three-particle basis sets to uncover the spin dynamics of our model. 
        The natural choice are states that are aligned to the action of the $\hat{S}^{z}=\hat{S}^{z}_{1}\otimes\hat{S}^{z}_{2}\otimes\hat{S}^{z}_{3}$ operator, i.e. the Hamiltonian and density operator are represented in the product basis $\ket{s,m_{s}}=\ket{s_{1},m_{s_{1}}}\otimes(\ket{s_{2},m_{s_{2}}}\otimes\ket{s_{3},m_{s_{3}}})$, where $\vb{S}=\vb{S}_{1}+\vb{S}_{2}+\vb{S}_{3}$. 
        Because the $S_{2,3}$ particle states are anticipated to be correlated within a Bloch sphere representation, we designate a ``device'' basis which is given by $\ket{s_{1},m_{1}}\otimes\ket{s_{23},m_{23}}$. 
        In this representation, $\ket{s_{23},m_{23}} = \ket{s_{2},m_{2}}\otimes\ket{s_{3},m_{3}}$ is designated the ``coupled particle'' basis.
\section{Results}

    \subsection{Condition for DJ Resonance}
    Switching is a dynamic process in which the exchange coupling ($J$) and the onsite anisotropy ($D$) act as competing interactions. Maximum switching occurs when these two interactions are perfectly balanced, a condition we call the DJ resonance \cite{switzer2021}. Such resonance has been observed experimentally in an ultra-cold atom system \cite{chung2021}.
    
\begin{figure}
        \includegraphics[width=\columnwidth]{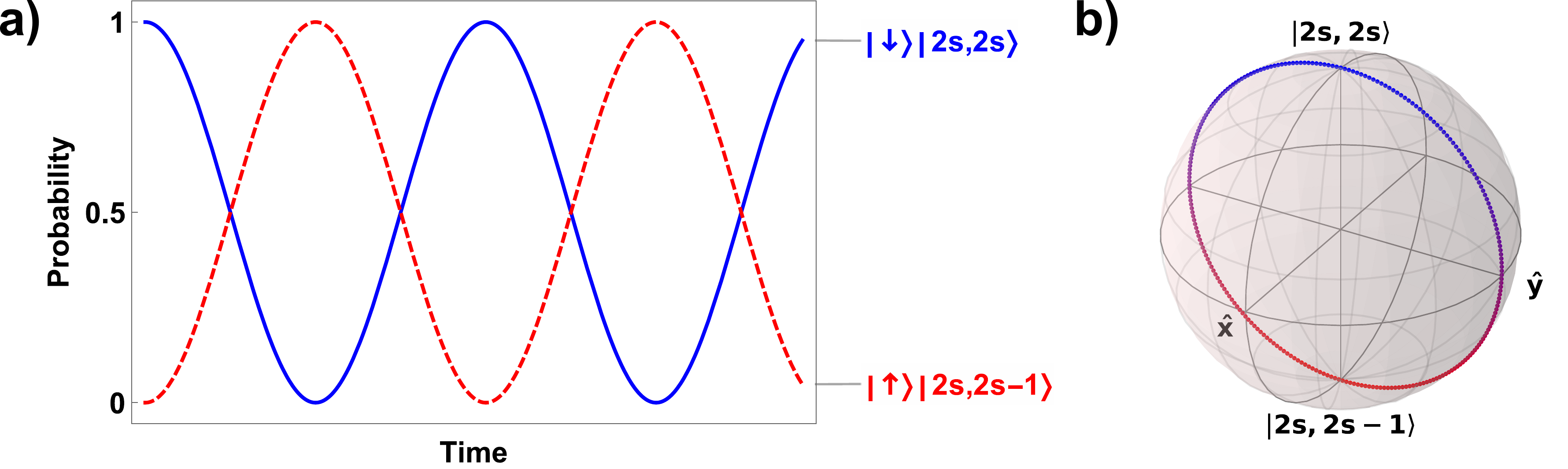}
        \colorcaption{\label{fig:djresonance}
        Representations of the DJ resonance for (a) the transition probability amplitudes between the $\ket{\downarrow}\ket{2s,2s}$ state (solid) and the $\ket{\uparrow}\ket{2s,2s-1}$ state (dashed), and (b) the corresponding Bloch sphere representation of the system.}
        \end{figure}

        We first demonstrate that for general $S_{2,3}>\tfrac{1}{2}$ and $S_{1}=\tfrac{1}{2}$, at least two DJ resonances exist. We begin by noting that for two particles with similar spin $s$, the following coupled particle basis states can be written in $\ket{s_{i},m_{s_{i}}}$ spin basis notation,
        \begin{align}
            &\ket{2s,2s}=\ket{s,s}\ket{s,s},\\
            &\ket{2s,2s-1}=\frac{1}{\sqrt{2}}\big(\ket{s,s}\ket{s,s-1}+\ket{s,s-1}\ket{s,s}\big).
        \end{align}
        We couple the $S_{1}$ particle to these coupled particle states, while maintaining spin angular momentum conservation. We choose, as will be apparent later, the two device states $\{\ket{\downarrow}\ket{2s,2s},\ket{\uparrow}\ket{2s,2s-1}\}$. Forming a subspace with these two states, one can easily see that the isotropic Heisenberg-like exchange Eq.~(\ref{eqn:hamiltonian-heisenberg-interaction}) and hopping Eq.~(\ref{eqn:hamiltonian-hopping}) terms are diagonal in the subspace, and can be neglected when considering the transition dynamics. We again do not restrict the range of the other exchange term, nor of the magnetic anisotropy. We thus concentrate on a reduced spin Hamiltonian of the form (dropping the $K$ subscript in the exchange term for notational ease),
        \begin{align}
        \label{eqn:reduced-hamiltonian}
            \mathcal{H}=J\hat{\vb{S}}_{1}\cdot\left(\hat{\vb{S}}_{2}+\hat{\vb{S}}_{3}\right)+D\left(\hat{S}^{z}_{2}\hat{S}^{z}_{2}+\hat{S}^{z}_{3}\hat{S}^{z}_{3}\right).
        \end{align}
        Applying this Hamiltonian to the chosen states within the device basis, we find that no other device states participate in the dynamics within this subspace. The Hamiltonian is then,
        \begin{align}
        \label{eqn:reduced-hamiltonian-device-basis-subblock}
        \mathcal{H}=\begin{pmatrix}
            2Ds^2-Js & J\sqrt{s} \\
            J\sqrt{s} & J\left(s-\frac{1}{2}\right)+D\left(s^{2}+\left(s-1\right)^2\right).
        \end{pmatrix}
        \end{align}
        Expressing it in a convenient form by removing a constant diagonal offset, we arrive at,
        \begin{align}
        \label{eqn:reduced-hamiltonian-device-basis-subblock-ppauli-matrices}
            \mathcal{H}=J\sqrt{s}\sigma_{x}+\left[D\left(s-\frac{1}{2}\right)-J\left(s-\frac{1}{4}\right)\right]\sigma_{z},
        \end{align}
        where $\sigma_{z}$ and $\sigma_{x}$ are the spin $\tfrac{1}{2}$ Pauli rotation matrices about the $\hat{z}$ and $\hat{x}$ axis, respectively.
        Applying the Rabi formula \cite{rabi37} to this Hamiltonian sub-block results in a Rabi frequency,
        \begin{align}
        \label{eqn:rabi-f-general-s}
            \Omega=\sqrt{\left(D\left(s-\frac{1}{2}\right)-J\left(s-\frac{1}{4}\right)\right)^{2}+\left(J\sqrt{s}\right)^{2}}.
        \end{align}
        If the system is prepared as a pure $\ket{\downarrow}\ket{2s,2s}$ state, the transition probability amplitude takes the form,
        \begin{align}
        \label{eqn:rabi-p-general-s}
            P(t)=\left(\frac{J\sqrt{s}}{\Omega}\right)^{2}\sin^{2}\left(\Omega t\right).
        \end{align}
        Resonance is achieved when Eq.~(\ref{eqn:rabi-p-general-s}) is maximized to unity. 
        This occurs when the rotation about the z-axis in this subspace is stopped, or when,
        \begin{align}
        \label{eqn:dj-resonance-general-s}
             J &= \frac{\left(s-\frac{1}{2}\right)}{\left(s-\frac{1}{4}\right)}D.
        \end{align}
        Performing a similar procedure on the subspace corresponding with the $\{\ket{\uparrow}\ket{2s,-2s},\ket{\downarrow}\ket{2s,-2s+1}\}$ states results in the same condition. 
        Thus there are at least two DJ resonances, one for each block described, for every value of $S_{2,3}>\tfrac{1}{2}$. A representation of the transition probability amplitudes and a corresponding Bloch sphere representation for any DJ resonance is given in Fig.~\ref{fig:djresonance}.
        
    \subsection{Generalized DJ Resonances}
        The analysis of the last section can be extended by considering a Hamiltonian with additional terms that contribute to an anisotropic spin response. Our new general spin Hamiltonian incorporates Eq.~(\ref{eqn:hamiltonian-heisenberg-interaction}), Eq.~(\ref{eqn:reduced-hamiltonian}), and an applied magnetic field along the same axis as the magnetic anisotropy Hamiltonian term. The general Hamiltonian is then,
        \begin{align}
        \label{eqn:generalized_djjb_hamiltonian}
        \mathcal{H}&=J\hat{\vb{S}}_{1}\cdot\left(\hat{\vb{S}}_{2}+\hat{\vb{S}}_{3}\right)+D\left(\hat{S}^{z}_{2}\hat{S}^{z}_{2}+\hat{S}^{z}_{3}\hat{S}^{z}_{3}\right)\\
        \nonumber
        &+J_{z}\hat{S}^{z}_{2}\hat{S}^{z}_{3}+J_{xy}\left(\hat{S}^{x}_{2}\hat{S}^{x}_{3}+\hat{S}^{y}_{2}\hat{S}^{y}_{3}\right)+\sum_{i=1}^{3}g_{i}\mu_{B}\hat{S}^{z}_{i}B_{0},
        \end{align}
        where $g_{i}$ is spin particle $i$'s g-factor, and $B_{0}$ is the applied static magnetic field strength for a field directed along the axis of particle 2 and 3's magnetic anisotropy. We assume that particle 2 and 3 have the same g-factor, i.e. $g_{2}=g_{3}=g_{23}$. Repeating the same procedure, one finds that when the Hamiltonian is expressed in the device basis, two blocks have the form,
        \begin{align}
        \label{eqn:generalized_djjb_hamiltonian_subblock_a}
            \mathcal{H}_{a}&=\Omega_{x}\sigma_{x}-\Omega_{z}\sigma_{z}-\frac{1}{2}\left(g_{1}-g_{23}\right)\mu_{B}B_{0}\sigma_{z}\\
        \label{eqn:generalized_djjb_hamiltonian_subblock_b}
            \mathcal{H}_{b}&=\Omega_{x}\sigma_{x}+\Omega_{z}\sigma_{z}-\frac{1}{2}\left(g_{1}-g_{23}\right)\mu_{B}B_{0}\sigma_{z},
        \end{align}
        where the ``a'' block corresponds with the dynamics of the $\{\ket{\downarrow}\ket{2s,2s},\ket{\uparrow}\ket{2s,2s-1}\}$ states, the ``b'' block corresponds with the dynamics of the $\{\ket{\downarrow}\ket{2s,-2s+1},\ket{\uparrow}\ket{2s,-2s}\}$ states, $\Omega_{x}=J\sqrt{s}$, and $\Omega_{z}=J\left(s-1/4\right)-D\left(s-1/2\right)+\left(J_{xy}-J_{z}\right)s/2$. Comparing the two blocks, one sees that the ``a'' block rotates an appropriate Bloch vector in that space clockwise about the $z$-axis. The applied magnetic field acts in concert with this rotation. Conversely, the ``b'' block rotates a corresponding Bloch vector counter-clockwise, with the applied magnetic field acting in competition with this rotation. This changes the resonance conditions, leading to two generalized DJ resonances,
        \begin{align}
            \nonumber
            J_{a}&=D\frac{\left(s-\frac{1}{2}\right)}{\left(s-\frac{1}{4}\right)}+\frac{1}{2}\frac{s}{\left(s-\frac{1}{4}\right)}\left(J_{z}-J_{xy}\right)\\
        \label{eqn:generalized_djjb_resonance_subblock_a}    
            &\;\;\;\;\;-\frac{1}{2}\frac{\mu_{B}B_{0}}{\left(s-\frac{1}{4}\right)}\left(g_{1}-g_{23}\right),\\
            \nonumber
            J_{b}&=D\frac{\left(s-\frac{1}{2}\right)}{\left(s-\frac{1}{4}\right)}+\frac{1}{2}\frac{s}{\left(s-\frac{1}{4}\right)}\left(J_{z}-J_{xy}\right)\\
        \label{eqn:generalized_djjb_resonance_subblock_b}
            &\;\;\;\;\;+\frac{1}{2}\frac{\mu_{B}B_{0}}{\left(s-\frac{1}{4}\right)}\left(g_{1}-g_{23}\right)
        \end{align}
        We next analyze the eigenvectors of the simpler Hamiltonian in Eq.~(\ref{eqn:reduced-hamiltonian}) for the two two-dimensional blocks considered so far in order to uncover the physical meaning behind these resonance conditions.
    \subsection{Physical Meaning of the DJ Resonance}
        We begin by examining which device basis states that participate in the block given in Eq.~(\ref{eqn:reduced-hamiltonian-device-basis-subblock-ppauli-matrices}) are energetically preferred. In the following discussion, we neglect the $s=\tfrac{1}{2}$ case, as we give results for that case in another section. Within that subspace, if $J=0$, and the magnetic anisotropy axis is hard, i.e. $D>0$, the $\ket{\uparrow}\ket{2s,2s-1}$ state is preferred. If the magnetic anisotropy axis is easy, i.e. $D<0$, the $\ket{\downarrow}\ket{2s,2s}$ state is preferred. A proof of this is easily shown by examining the eigensystem of the block when $J=0$. The eigenvalues are $\mp D\left(s-\frac{1}{2}\right)$ which are matched to the eigenvectors $\ket{\uparrow}\ket{2s,2s-1}$ and $\ket{\downarrow}\ket{2s,2s}$, respectively. 
        
        The opposite preference occurs when $D=0$ and $J \neq 0$. If $J$ represents antiferromagnetic exchange coupling (i.e. when $J$ is positive in our sign convention), then the $\ket{\downarrow}\ket{2s,2s}$ is favored. If $J$ represents ferromagnetic exchange coupling (i.e. $J$ is negative), then the $\ket{\uparrow}\ket{2s,2s-1}$ is favored. Thus when inspecting the regime of the DJ resonance (when $J$ and $D$ have the same sign and $s>\tfrac{1}{2}$), one can see that the magnetic anisotropy and exchange terms of the total Hamiltonian act in competition.
        
        Next, we examine the eigensystem of the block for relevant conditions for $s>\tfrac{1}{2}$. The eigenvalues are $\alpha=\mp\sqrt{\beta^{2}+\left(J\sqrt{s}\right)^{2}}$, where $\beta\equiv D\left(s-\tfrac{1}{2}\right)-J\left(s-\tfrac{1}{4}\right)$. The corresponding eigenvectors are,
        \begin{align}
            \ket{\phi_{1}}=&\rightarrow A \left(\frac{\beta-\alpha}{J\sqrt{s}}\right)\ket{\downarrow}\ket{2s,2s}+A\ket{\uparrow}\ket{2s,2s-1},\\
            \ket{\phi_{2}}=&\rightarrow A \left(\frac{\beta+\alpha}{J\sqrt{s}}\right)\ket{\downarrow}\ket{2s,2s}+A\ket{\uparrow}\ket{2s,2s-1},
        \end{align}
        where $A$ is a normalization constant. It is s apparent that the $\ket{\phi_{1}}$ eigenstate is always preferred except when $\alpha=0$. The latter condition is never satisfied for $J\neq 0$, i.e. there are no condition in which the $\ket{\phi_{2}}$ state is energetically favorable.
        
\begin{figure}
        \includegraphics[width=\columnwidth]{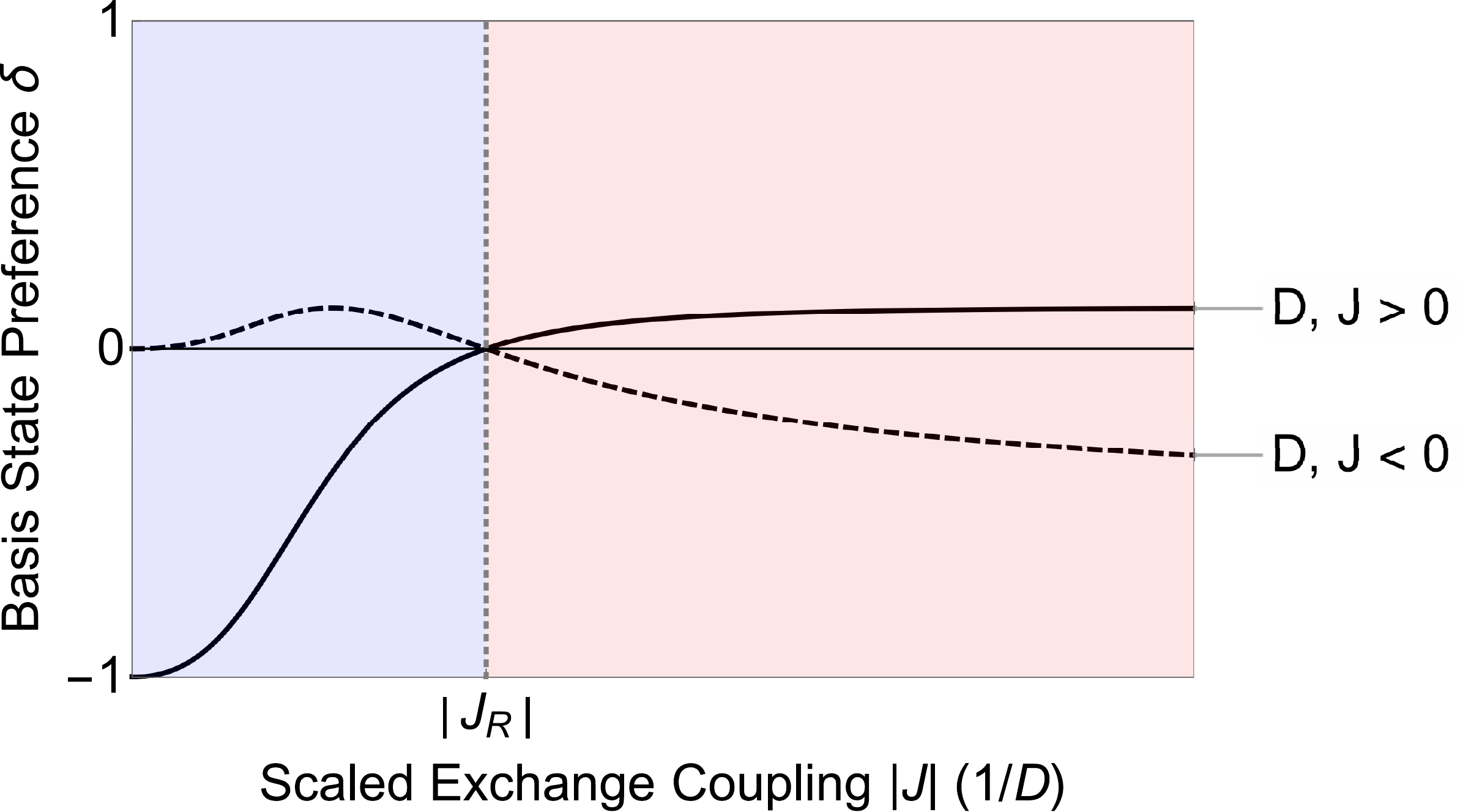}
        \colorcaption{\label{fig:eigenvectors}
        Representation of basis set mixture in the energetically-favored eigenvector $\ket{\phi_{1}}$ when $D$ and $J>0$ (solid) and when $D$ and $J<0$ (dashed). The difference between the modulus squared coefficients $\delta\equiv\abs{c_{1}}^{2}-\abs{c_{2}}^{2}$ is plotted for $s=1$, where $c_{1}$ is the coefficient of the $\ket{\downarrow}\ket{2s,2s}$ basis state within the $\ket{\phi_{1}}$ eigenvector, and $c_{2}$ the coefficient of the $\ket{\uparrow}\ket{2s,2s-1}$ state. Basis state mixtures that are favorable to magnetic anisotropy are shaded blue, while those that favor exchange coupling are shaded red.}
        \end{figure}

        Armed with the energetically favorable eigenvector, we now uncover the balance of device basis states around the DJ resonance. This can be done by examining the relative proportions of the two device basis states within the $\ket{\phi_{1}}$ eigenvector. For example, if the eigenvector contains a larger proportion of the $\ket{\downarrow}\ket{2s,2s}$ basis state for positive $D$ and $J$, then one can infer from the prior analysis that the exchange coupling Hamiltonian term plays a stronger role in the dynamics than the magnetic anisotropy term. The test is then to compare the modulus squared values of the components of the eigenvector. When both components are balanced (and noting that both components are always real), the condition is,
        \begin{align}
            \left(J\sqrt{s}\right)^{2}=\left(\beta-\alpha\right)^{2}.
        \end{align}
        This condition holds exactly at the DJ resonance given in Eq.~(\ref{eqn:dj-resonance-general-s}). We repeat this analysis for off resonance conditions, i.e when $J=J_{\text{R}}+\epsilon$ for appropriate $\epsilon$. We also enforce conditions that lie within the DJ resonance regime, i.e. $J \in ( 0,\infty ) $ when $D>0$ and $J \in (-\infty, 0 )$ when $D<0$. A representation of these conditions, and the results, are shown in Fig.~\ref{fig:eigenvectors}. 
        We find that when $\abs{J}<\abs{J_{\text{R}}}$, there is a higher proportion of the basis states within the eigenvector that are energetically favorable to magnetic anisotropy. Conversely, when $\abs{J}>\abs{J_{\text{R}}}$, there is a lower proportion of those same basis states in the eigenvector. Thus the DJ resonance is the inflection point in which the action of the two terms in the total Hamiltonian are in perfect balance.
        We next explore the $S_{2,3}=\tfrac{1}{2}$ model, in light of the condition that Eq.~(\ref{eqn:dj-resonance-general-s}) takes at that value of $s$, and compare to the $S_{2,3}=1$ model.

    \subsection{Coupled Particles with \texorpdfstring{$S_{2,3}=\tfrac{1}{2}$}{TEXT} Model}
        For the $S_{2,3}=\frac{1}{2}$ model, when the Hamiltonian of Eq.~(\ref{eqn:hamiltonian-total}) is expressed in the product basis, the two $s=\frac{1}{2}$ subspaces are coupled to each other by anisotropic application of the exchange term $J_{\text{K}}$, while the $s=\frac{3}{2}$ subspace is diagonal and remains uncoupled. 
        The Heisenberg-like exchange, given in Eq.~(\ref{eqn:hamiltonian-heisenberg-interaction}), is diagonal in both the product and device basis if $J_{z}=J_{xy}=J_{\text{H}}$, i.e. the interaction is isotropic between particle 2 and 3. While it's important to consider the situation in which the Heisenberg-like exchange is applied anisotropically (i.e. $J_{z}\neq J_{xy}$), we first consider the isotropic case. The tunneling Hamiltonian in Eq.~(\ref{eqn:hamiltonian-hopping}) is also diagonal in both the product and device basis. 
        In the device basis, the tunneling Hamiltonian carries the same eigenvalue across all $s_{23}$ subspaces in the device basis, and the isotropic Heisenberg exchange Hamiltonian has the same eigenvalue in each subspace.
        As a result, the total Hamiltonian in the product basis takes the form,
        \begin{align}
        \label{eqn:hamiltonian-productbasis-s23-1d2}
            \begin{pmatrix}
                \begin{bmatrix}
                    \mathcal{H}^{0}_{3/2}
                \end{bmatrix} &  &  \\
                 & \begin{bmatrix}
                    \mathcal{H}^{1}_{1/2} 
                 \end{bmatrix} &  \\
                 &  & \begin{bmatrix} 
                 \mathcal{H}^{2}_{1/2}
                 \end{bmatrix} \\
            \end{pmatrix},
        \end{align}
        where $\mathcal{H}^{n}_{1/2}$ refers to the Hamiltonian block that corresponds with the $n$'th interaction of the two spin $s=\frac{1}{2}$ subspaces, and $\mathcal{H}^{0}_{3/2}$ is the diagonal $s=\frac{3}{2}$ subspace. 
        When the Hamiltonian is expressed in the device basis, the anisotropic application of the exchange term couples the different $s_{2,3}$ subspaces. 
        This leads to a block Hamiltonian of the form,
        \begin{align}
        \label{eqn:hamiltonian-devicebasis-s23-1d2}
            \begin{pmatrix}
                \begin{bmatrix}
                    \mathcal{H}^{0}_{\pm3/2}
                \end{bmatrix} &  &  \\
                 & \begin{bmatrix}
                    \mathcal{H}_{1/2} 
                 \end{bmatrix} &  \\
                 &  & \begin{bmatrix} 
                 \mathcal{H}_{-1/2}
                 \end{bmatrix} \\
            \end{pmatrix},
        \end{align}
        where $\mathcal{H}_{m}$ refers to a 3 dimensional Hamiltonian block that corresponds with the $m$ subspace, and $\mathcal{H}^{0}$ refers to a 2 dimensional diagonal block of spin $s_{23} = 1$. 
        
        We further explore the $m=\pm1/2$ blocks corresponding with the $\ket{m_{1}}\ket{s_{23},m_{23}}=\ket{\uparrow}\ket{1,0},\;\ket{\uparrow}\ket{0,0},\;\ket{\downarrow}\ket{1,1}$ and $\ket{\uparrow}\ket{1,-1},\;\ket{\downarrow}\ket{1,0},\;\ket{\downarrow}\ket{0,0}$ states, respectively. 
        They take the form (with a common $t+t^{*}+\frac{1}{4}J_{\text{H}}$ removed from the diagonal),
        \begin{align}
        \label{eqn:hamiltonian-subblock-1d2-devicebasis}
            \mathcal{H}_{1/2}&=\frac{1}{4}
            \begin{pmatrix}
                0 & \Delta_{\text{K}} & \sqrt{2}\Sigma_{\text{K}}\\
                \Delta_{\text{K}} & -4J_{H} & -\sqrt{2}\Delta_{\text{K}}\\
                \sqrt{2}\Sigma_{\text{K}} & -\sqrt{2}\Delta_{\text{K}} & -\Sigma_{\text{K}}
            \end{pmatrix},\\
            \mathcal{H}_{-1/2}&=\frac{1}{4}
            \begin{pmatrix}
                -\Sigma_{\text{K}} & \sqrt{2}\Sigma_{\text{K}} & \sqrt{2}\Delta_{\text{K}} \\
                \sqrt{2}\Sigma_{\text{K}} & 0 & -\Delta_{\text{K}} \\
                \sqrt{2}\Delta_{\text{K}} & -\Delta_{\text{K}} & -4J_{H}
            \end{pmatrix},
        \end{align}
        where  $\Delta_{\text{K}}\equiv J_{\text{K2}}-J_{\text{K3}}$ and $\Sigma_{\text{K}}\equiv J_{\text{K2}}+J_{\text{K3}} \equiv 2 J_{\text{K}}$.
        When the exchange term is isotropically applied, i.e. $\Delta_{\text{K}}=0$, the corresponding $\ket{0,0}$ states no longer interact with the others. The Hamiltonian blocks then become,
        \begin{align}
        \label{eqn:hamiltonian-isokondo-subblock-s0.5-1d2-devicebasis}
            \mathcal{H}^{eff}_{\pm1/2}&=
            \pm\frac{1}{4}J_{\text{K}}
            \sigma_{z}+
            \frac{1}{\sqrt{2}}J_{\text{K}}
            \sigma_{x}.
        \end{align}
        
\begin{figure}
        \includegraphics[width=\columnwidth]{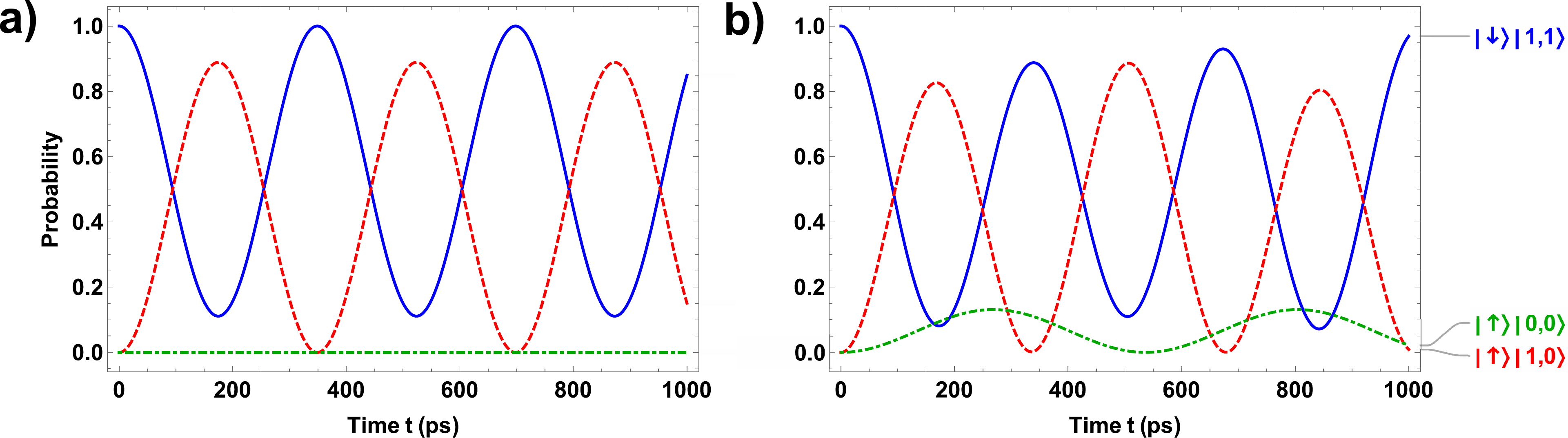}
        \colorcaption{\label{fig:spin0.5anivsiso}
        Transition probability amplitudes for the for the $S_{2,3}=\tfrac{1}{2}$ model, for the $\ket{\downarrow}\ket{1,1}$ (solid), $\ket{\uparrow}\ket{1,0}$ (dashed), and $\ket{\uparrow}\ket{0,0}$ (dot-dashed) states as a function of time. 
        The exchange strength is (a) isotropically-applied $J_{\text{K}}$ and (b) anisotropically-applied $J_{\text{K}}$ (i.e. $J_{\text{K2}} \ne J_{\text{K3}}$). 
        The system is initially prepared as a pure $\ket{\downarrow}\ket{1,1}$ state. 
        In units of $\text{cm}^{-1}$, exchange parameters are (a) $J_{\text{K2}}=J_{\text{K3}}=-0.40$ and (b) $\Delta_{\text{K}}/J_{\text{K}}=0.5$. 
        The shared parameters are: $J_{\text{H}}=-0.05$ and $t=0.05$.
        }
        \end{figure}

        Applying the Rabi formula to the case in which the system is initially prepared in either the pure $\ket{\downarrow}\ket{1,1}$ or $\ket{\uparrow}\ket{1,-1}$ state, we find,
        \begin{align}
        \label{eqn:rabi-p-1d2}
            P(t) &= \frac{8}{9}\sin^{2}(\Omega t),
        \end{align}
        with Rabi frequency,
        \begin{align}
        \label{eqn:rabi-f-1d2}
            \Omega &= \frac{3}{4}\abs{J_{\text{K}}}.
        \end{align}
        In this case, and as shown in Fig.~\ref{fig:spin0.5anivsiso}, there are no resonance conditions in which the maximum transition probability amplitude is unity, even when considering anisotropic application of the exchange coupling. 
        
        To check for scenarios in which a maximum transition probability of unity can be achieved in the $S_{2,3}=\tfrac{1}{2}$ model, we explore applications of spin filtering on the system. To do this, first an appropriate change of basis of the device states is needed. 
        The states, originally expressed in the $\ket{s_{1},m_{s_{1}}}\otimes\ket{s_{2,3},m_{s_{2,3}}}$, are changed by transforming $\ket{s_{1},m_{s_{1}}}$ states to $\ket{\theta,\phi}$ spinor states. 
        In general, the spinor takes the form,
        \begin{align}
        \label{eqn:spinor}
            \chi =
            \begin{pmatrix}
                \cos{\frac{\theta}{2}}e^{-i\frac{\phi}{2}} \\
                \sin{\frac{\theta}{2}}e^{i\frac{\phi}{2}}
            \end{pmatrix},
        \end{align}
        where $\theta$ is the polar angle and $\phi$ is the azimuthal angle. 
        Because the reduced Hamiltonian blocks will no longer be two dimensional, we numerically solve Eq.~(\ref{eqn:time-dependent-density-matrix}) to obtain the relative transition probabilities from one state to the next. 
        
\begin{figure}
        \includegraphics[width=\columnwidth]{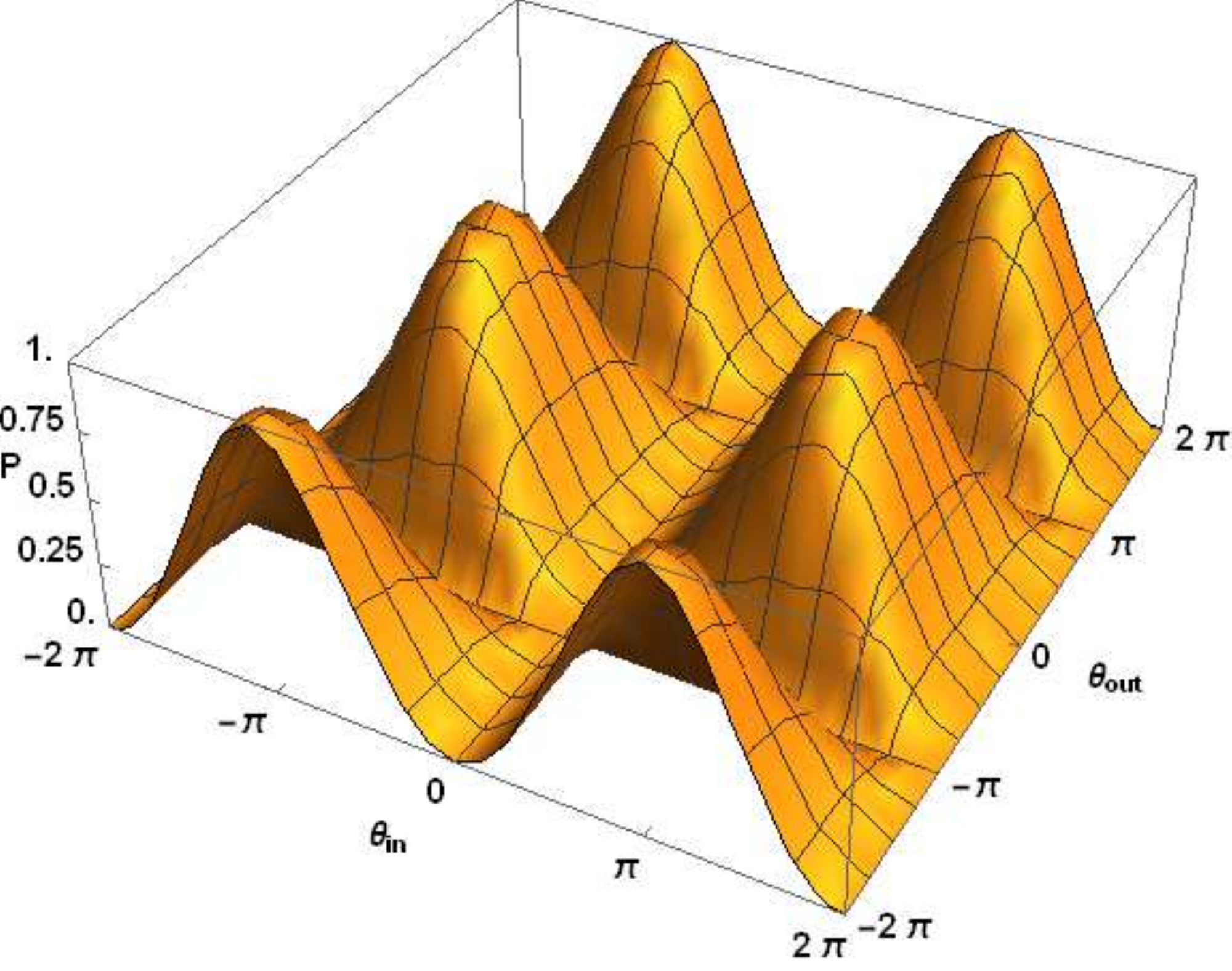}
        \caption{\label{fig:filter-search}
        Relative transition probability ($z$ axis) of the $\ket{\theta,\phi}\ket{s_{23},m_{23}}=\ket{\theta_{\text{in}},0}\ket{1,1}\rightarrow\ket{\theta_{\text{out}},0}\ket{1,0}$ with respect to the total probability of finding the system with the chosen $S_{1}$ particle's measured polar spin orientation ($y$ axis), as a function of the chosen $S_{1}$ particle's prepared polar spin orientation ($x$ axis).
        The snapshot of probabilities was calculated for $t = 133$ ps. 
        In units of $\text{cm}^{-1}$, the parameters are $J_{\text{K2}}=J_{\text{K3}}=-0.40$, $J_{\text{H}}=-0.05$, $t=0.05$.}
        \end{figure}

\begin{figure}
        \includegraphics[width=\columnwidth]{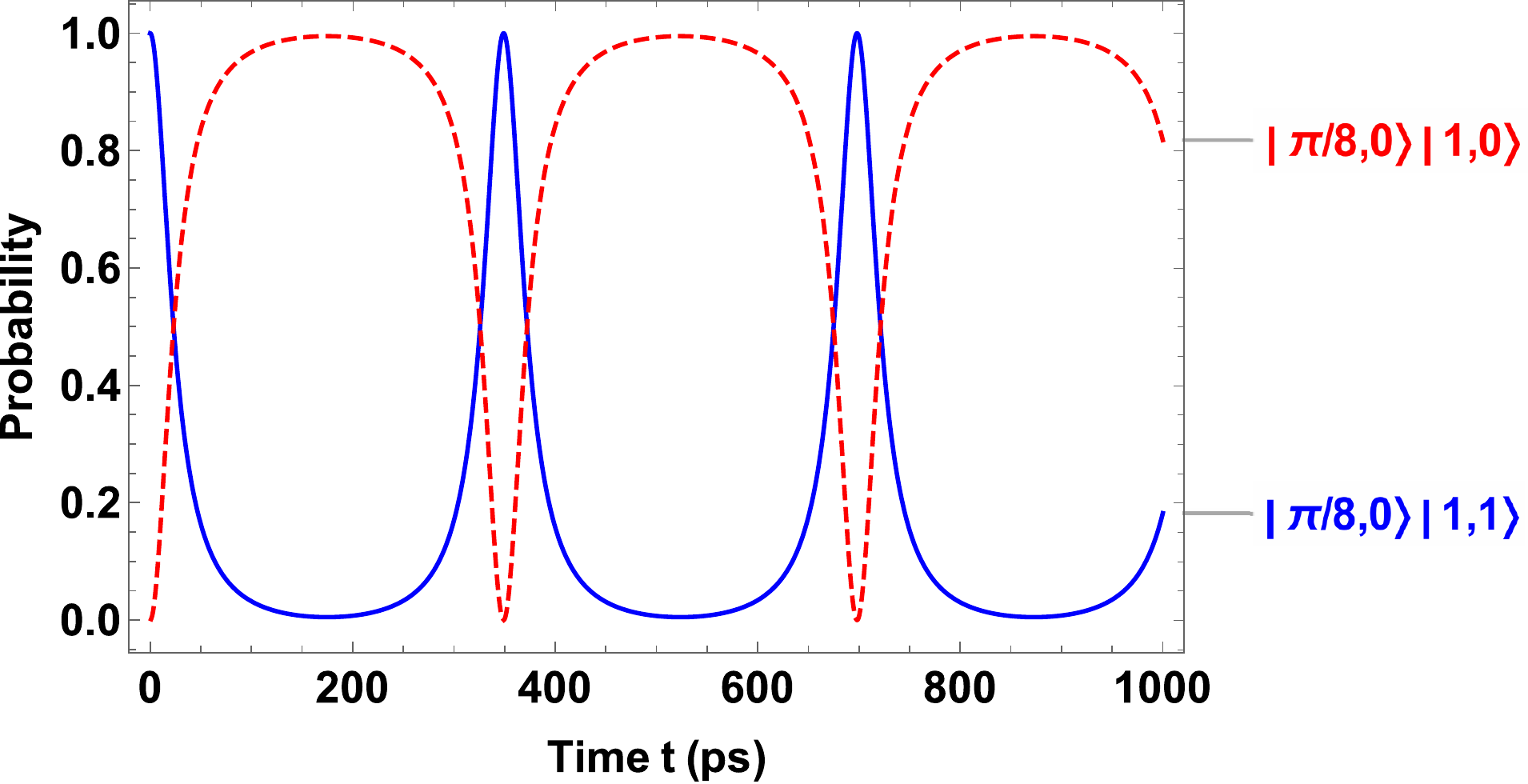}
        \caption{\label{fig:filtered-transition}
        Relative transition probability amplitudes in the $S_{\text{2,3}}=\frac{1}{2}$ model, for the $\ket{\pi/8}\ket{1,1}$ (solid) and $\ket{\pi/8}\ket{1,0}$ (dashed) states, as a function of time. 
        The system is initially prepared in the pure $\ket{\pi,0}\ket{1,1}$ state, with the same parameter state as Fig.~\ref{fig:filter-search}.}
        \end{figure}

        We choose to monitor those states that are coupled in the Hamiltonian and correspond with transitions between non-entangled to maximally entangled states. 
        The transition $\ket{1,1}\rightarrow\ket{1,0}$ fits this description, as the device-to-spin correspondence is,
        \begin{align}
        \label{eqn:device2site-s23-1d2-transition}
            \ket{1,1} &= \ket{\uparrow}\ket{\uparrow},\\
            \ket{1,0} &= \frac{1}{\sqrt{2}}\left(\ket{\uparrow} \ket{\downarrow}+\ket{\downarrow}\ket{\uparrow}\right).
        \end{align}
        Fig.~\ref{fig:filter-search} investigates the orientations of particle 1's spin that will lead to maximum probability between these two states at a given snapshot of the Rabi-like oscillations, with the azimuthal orientation angle chosen to be $0$. 
        Peaks of transition probability are found for several combinations of preparation and measurement orientations. 
        Taking Fig.~\ref{fig:filter-search} as guidance to explore a higher transition probability scenario for a given spin filter, particle 1's prepared polar angle is chosen to be $\theta_{\text{in}}=\pi$ and the measured polar angle to be $\theta_{\text{out}}=\frac{\pi}{8}$. 
        The resulting transition probability oscillations are shown in Fig.~\ref{fig:filtered-transition}. 
        It is clear from this result that the maximum amplitude has indeed been raised from $8/9$, calculated to be $P=0.995$.
        Fig.~\ref{fig:filtered-transition} demonstrates the same feature of the spin dynamics for the three-particle system as found in Ref.~\cite{switzer2021}, namely that a single measurement of particle 1's spin orientation determines the entanglement state of particle 2 and 3. 
        This demonstrates the read out of the entanglement state if the measurement of the $S_{1}$ spin polarization is taken at any general time $t$. 
        This also demonstrates preparation of the entanglement state if the $S_{1}$ spin polarization is measured at a specific time $t$ corresponding with a peak in the Rabi oscillation.
        
        Finally, another method may be used to prepare the entanglement state. Looking to Eq.~(\ref{eqn:generalized_djjb_resonance_subblock_a}) and Eq.~(\ref{eqn:generalized_djjb_resonance_subblock_b}), the resonance condition for $S_{2,3}=\tfrac{1}{2}$ without an applied magnetic field is,
        \begin{align}
            J_{\text{K}} = J_{z}-J_{xy}.
        \end{align}
        When this resonance condition (which we designate as the ``JJ resonance'') is applied to the system considered in Fig.~\ref{fig:spin0.5anivsiso}(a), the maximum transition probability amplitude is unity. Thus while the $S_{2,3}=\tfrac{1}{2}$ model does not contain a DJ resonance as indicated in Eq.~(\ref{eqn:dj-resonance-general-s}), a JJ resonance could be used to achieve the same transition probability behavior.
        As we will next show, and in comparison to the $S_{2,3}=\tfrac{1}{2}$ model, this resonance switching mechanism is maximized in the $S_{2,3}=1$ model without the use of spin filtering mechanisms or anisotropy of the exchange interaction between particle 2 and 3.

    \subsection{Comparison to the Coupled Particles \texorpdfstring{$S_{2,3}=1$}{TEXT} Model}
        We next compare the $S_{2,3}=1$ model as explored in \cite{switzer2021} by identifying the role of each Hamiltonian term given in Eq.~(\ref{eqn:hamiltonian-total}) in that model.
        Similar to the $S_{2,3}=\tfrac{1}{2}$ model, the Heisenberg-like exchange and tunneling Hamiltonians are diagonal in the device basis. Because the magnetic anisotropy term is no longer diagonal like in the $S_{2,3}=\tfrac{1}{2}$ model, the $s=2$ and $s=0$ subspaces are connected.
        As shown in Ref.~\cite{switzer2021}, because the magnetic anisotropy and exchange Hamiltonians have off-diagonal terms in similar $m$ subspaces, there is a complicated interplay of these interactions that influence the transition dynamics of systems prepared for a given value of $m$.
        
\begin{figure}
        \includegraphics[width=\columnwidth]{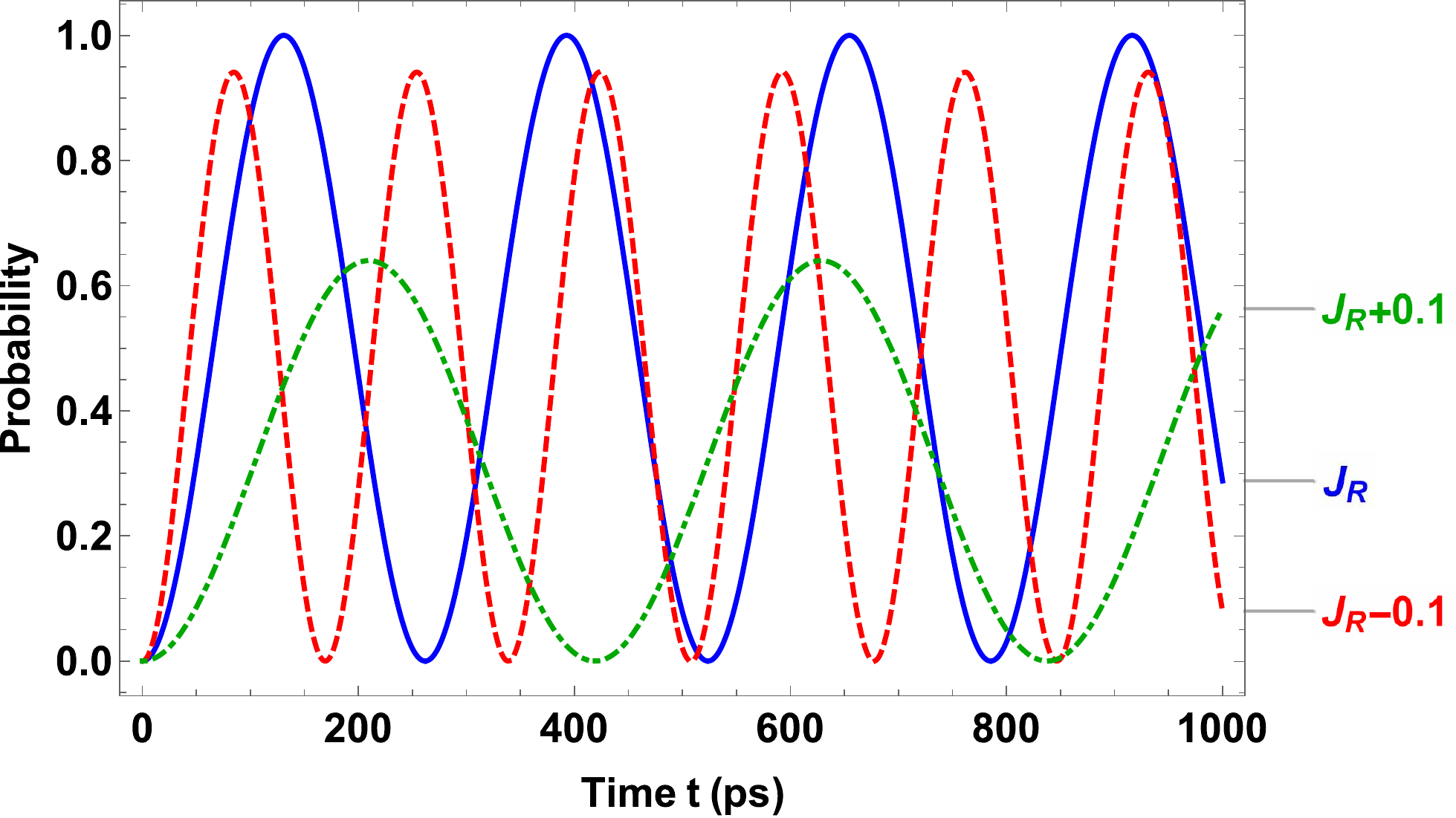}
        \colorcaption{\label{fig:spin1range}
        Transition probability amplitudes as a function of time for the $S_{2,3}=1$ model, for the $\ket{m_{1}}\ket{s_{23},m_{23}}=\ket{\uparrow}\ket{2,1}$ state when the initial state is prepared as a pure $\ket{\downarrow}\ket{2,2}$ state. 
        Several values of the isotropically-applied $J_{\text{K}}$ (i.e. $J_{\text{K2}}=J_{\text{K3}}$) are given. 
        The exchange strengths correspond with the DJ resonance condition of $J_{K}=\frac{2}{3}D$ (solid) and off resonance by $\pm0.1$ (dot-dashed and dashed, respectively). 
        In units of $\text{cm}^{-1}$, the parameters are: $J_{\text{H}}=-0.05$, $J_{\text{R}}=-0.40$, $D=-0.60$, $t=0.05$.
        }
        \end{figure}

        We next summarize the blocks of the $S_{2,3}=1$ model Hamiltonian. The $m=\tfrac{3}{2}$ subspace, which corresponds with the dynamics of the $\ket{m_{1}}\ket{s_{23},m_{23}}=\left\{\ket{\uparrow}\ket{2,1},\ket{\uparrow}\ket{1,1},\ket{\downarrow}\ket{2,2}\right\}$ states, takes the form (after removing a common $t+t^{*}+J_{\text{H}}+D+\frac{1}{4}\Sigma_{\text{K}}$ from the diagonal),
        \begin{align}
        \label{eqn:hamiltonian-subblock-3d2-devicebasis}
            \mathcal{H}_{3/2}&=\frac{1}{4}
            \begin{pmatrix}
                0 & \Delta_{\text{K}} & 2\Sigma_{\text{K}}\\
                \Delta_{\text{K}} & -8J_{H} & -2\Delta_{\text{K}}\\
                2\Sigma_{\text{K}} & -2\Delta_{\text{K}} & -3\Sigma_{\text{K}}+4D
            \end{pmatrix}.
        \end{align}
        Similarly, the $m=-\tfrac{3}{2}$ block (corresponding with the dynamics of the $\left\{\ket{\uparrow}\ket{2,-2},\ket{\downarrow}\ket{2,-1},\ket{\downarrow}\ket{1,-1}\right\}$ states) with the same diagonal values removed has the form,
        \begin{align}
        \label{eqn:hamiltonian-subblock-n3d2-devicebasis}
            \mathcal{H}_{-3/2}&=\frac{1}{4}
            \begin{pmatrix}
                -3\Sigma_{\text{K}}+4D & 2\Sigma_{\text{K}} & 2\Delta_{\text{K}}\\
                2\Sigma_{\text{K}} & 0 & -\Delta_{\text{K}} \\
                2\Delta_{\text{K}} & -\Delta_{\text{K}} & -8J_{H}
            \end{pmatrix}.
        \end{align}
        When the exchange coupling is instead isotropically applied, and similar to the $S_{2,3}=\tfrac{1}{2}$ model, the $\ket{\uparrow}\ket{1,1}$ state in Eq.~(\ref{eqn:hamiltonian-subblock-3d2-devicebasis}) and the $\ket{\downarrow}\ket{1,1}$ state in Eq.~(\ref{eqn:hamiltonian-subblock-n3d2-devicebasis}) are no longer coupled to the other states within their respective block Hamiltonians. 
        For the other two states in the $m=\pm\tfrac{3}{2}$ blocks, the effective Hamiltonians (with the appropriate diagonal entries removed) become, 
        \begin{align}
        \label{eqn:hamiltonian-isokondo-subblock-3d2-devicebasis}
            \mathcal{H}^{eff}_{\pm3/2}&=
            \mp\frac{1}{2}\left(D-\frac{3}{2}J_{\text{K}}\right)
            \sigma_{z}+
            J_{\text{K}}
            \sigma_{x}.
        \end{align}
        While the anisotropic case of exchange coupling is more complicated for the $m=\pm\tfrac{1}{2}$ subspaces, a similar reduction in the corresponding block Hamiltonians is found when the exchange coupling is applied isotropically. 
        Repeating the same procedure, we find,
        \begin{align}
        \label{eqn:hamiltonian-isokondo-subblock-1d2-devicebasis}
            \mathcal{H}^{eff}_{\pm1/2}&=
            \pm\frac{1}{2}\left(D+\frac{1}{2}J_{\text{K}}\right)
            \sigma_{z}+
            \frac{1}{\sqrt{2}}J_{\text{K}}
            \sigma_{x}.
        \end{align}
        Thus the same procedure to analyze the Rabi frequency of the $m$ subblocks can be used. As shown in Fig.~\ref{fig:spin1range}, if $J_{\text{K}}$ is chosen to be minimize the Rabi frequency, e.g. when $J_{\text{K}}=\tfrac{2}{3}D$ for $m=\pm\tfrac{3}{2}$, the maximum transition probability amplitude is unity. 
        When $J_{\text{K}}$ is chosen off of the resonance condition, the denominator in the Rabi formula increases and the maximum transition probability amplitude decreases.
        
        
\section{Discussion and Summary}
    
        \subsection{Control of Degree of Entanglement and Vectors on the Bloch Sphere}
        The main result of this work, demonstrated effectively through Eq.~(\ref{eqn:dj-resonance-general-s}), is that for any value of $S_{2,3}>\tfrac{1}{2}$, with no restriction on half-integer versus integer spin, at least two DJ resonances exist.
        Taking the classical limit as $s\rightarrow\infty$, the resonance condition tends towards $J_{K}=D$. 
        As is shown in Table \ref{tab:resonance-s1-s0.5}, Fig.~\ref{fig:filtered-transition}, and Fig.~\ref{fig:spin1range}, both the $S_{2,3}=\frac{1}{2}$ and $S_{2,3}=1$ models demonstrate preparation and measurement of the coupled particles' degree of entanglement by appropriate measurement of the $S_{1}=\frac{1}{2}$ particle.
        \begin{table}
        \caption{\label{tab:resonance-s1-s0.5}
        Pure state transitions for the $S_{2,3}=1$ and $S_{2,3}=1/2$ model, where $J_{\text{R}}$ is the condition on $J_{\text{K}}$ to reach resonance, $P$ is the maximum transition probability amplitude, and $\Omega$ is the Rabi frequency for this maximum amplitude.}
        \begin{ruledtabular}
            \begin{tabular}{lllll}
            $S_{2,3}$ & State Transitions & $J_{\text{R}}$ & $P$ & $\Omega$ \\
            
            $\frac{1}{2}$ & $\ket{\uparrow}\ket{1,0},\;\ket{\downarrow}\ket{1,+1}$ & - & $\frac{8}{9}$ & $\frac{3}{4}\abs{J_{\text{K}}}$\\
            
            $\frac{1}{2}$ & $\ket{\downarrow}\ket{1,0},\;\ket{\uparrow}\ket{1,-1}$ & - & $\frac{8}{9}$ & $\frac{3}{4}\abs{J_{\text{K}}}$\\
            
            $1$ & $\ket{\uparrow}\ket{2,+1},\;\ket{\downarrow}\ket{2,+2}$ & $\frac{2}{3}D$ & $1$ & $\frac{2}{3}\abs{D}$\\
            
            $1$ & $\ket{\uparrow}\ket{2,-2},\;\ket{\downarrow}\ket{2,-1}$ & $\frac{2}{3}D$ & $1$ & $\frac{2}{3}\abs{D}$\\
            
            $1$ & $\ket{\uparrow}\ket{1,0},\;\ket{\downarrow}\ket{1,+1}$ & $-2D$ & $1$ & $\sqrt{2}\abs{D}$\\
            
            $1$ & $\ket{\uparrow}\ket{1,-1},\;\ket{\downarrow}\ket{1,0}$ & $-2D$ & $1$ & $\sqrt{2}\abs{D}$
            
            \end{tabular}
            \end{ruledtabular}
        \end{table}
        
        We find that the physical interplay between the exchange and anisotropy terms in the Hamiltonian are vital for realizing the DJ resonance. As shown in the $S_{2,3}=\frac{1}{2}$ model, the lack of anisotropy in particle 2 and 3 result in no DJ resonances. We also find that at each DJ resonance, the system has achieved a perfect balance of states that balance the effect of the exchange coupling and magnetic anisotropy terms in the total Hamiltonian. In other words, if $\abs{J_{\text{K}}}=\abs{J_{R}}$, where $J_{R}$ is a DJ resonance, the most energetically favorable eigenvector of the Hamiltonian sub block contains an equal mixture of states that favor magnetic anisotropy and exchange coupling.
    
\begin{figure}
        \includegraphics[width=\columnwidth]{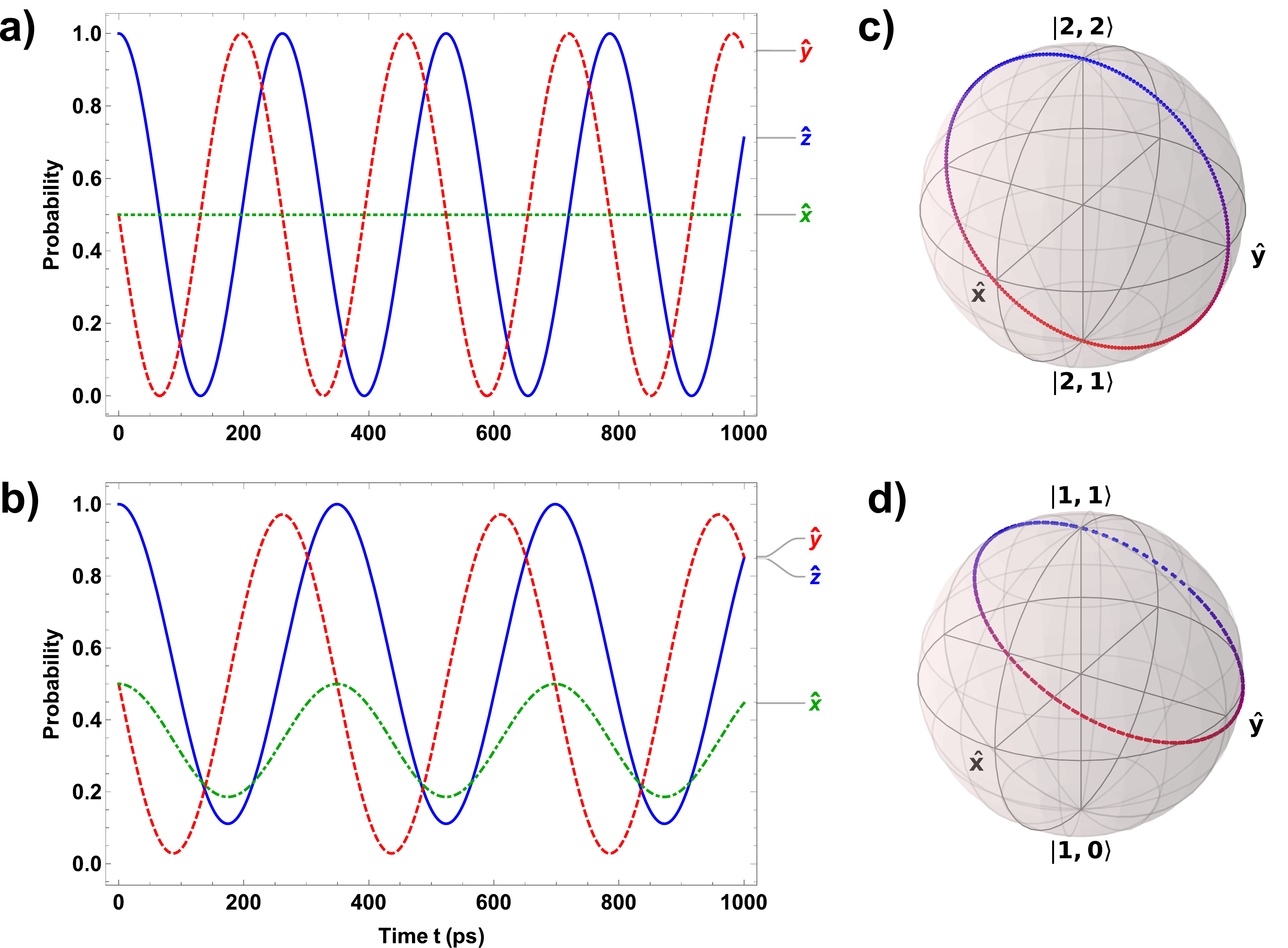}
        \colorcaption{\label{fig:resonancecomparison}
        Probability of measuring a state corresponding with the $\hat{x}$ (dotted), $\hat{y}$ (dashed), and $\hat{z}$ (solid) unit vectors as a function of time for the (a) $S_{2,3}=1$ and (b) $S_{2,3}=\tfrac{1}{2}$ models on the Bloch sphere as defined in (c) and (d), respectively. 
        The Bloch vector for (a)/(c) and (b)/(d) is initially prepared in the $\ket{\downarrow}\ket{2,2}$ and $\ket{\downarrow}\ket{1,1}$ states, respectively. 
        The corresponding Bloch sphere representation of the path traced by the Bloch vector for the (c) $S_{2,3}=1$ and (d) $S_{2,3}=\tfrac{1}{2}$ model is given for the same interval considered. 
        In units of $\text{cm}^{-1}$, the shared parameters are: $J_{\text{K2}}=J_{\text{K3}}=-0.40$, $J_{\text{H}}=-0.05$, $t=0.05$. The $S_{2,3}=1$ model additionally has $D=-0.60$.
        }
        \end{figure}

        For the $S_{2,3}=\frac{1}{2}$ model, the non-maximal probability amplitude indicates in the language of the Bloch sphere that the rotation is not about an axes solely on the azimuthal plane, but instead contains a rotation axis component in the polar plane, as shown in Fig.~\ref{fig:resonancecomparison}.
        By searching for appropriate spin filtering conditions, as shown in Fig.~\ref{fig:filter-search}, one could find rotations in the Bloch sphere that maximizes the amount of time spent near the non-entangled to entangled transition is possible.
        The DJ resonances as found in \cite{switzer2021}, however, is a feature for the $S_{2,3}>\tfrac{1}{2}$ model. 
        Rather than relying on either the use of extra spin filtering step to maximize transition probabilities, or the use of other anisotropies such as the anisotropic application of the Heisenberg-like term, the DJ resonance allows for full control of the Bloch sphere representation with appropriate control of the exchange coupling strength on and off resonance.
        
        Control of the rotation of a general Bloch vector within the appropriate Bloch sphere is important in QIS contexts. 
        Looking to Eq.~(\ref{eqn:generalized_djjb_hamiltonian_subblock_a}) and Eq.~(\ref{eqn:generalized_djjb_hamiltonian_subblock_b}), the exchange coupling $J$ gives one control over rotations of the Bloch vector about the $x$-axis, and the resonance condition allows control of rotations about the $z$-axis.
        With appropriate pulsing of the parameters, one can realize any point on the corresponding Bloch sphere. 
        It's precisely because of the resonance condition that all points on the sphere are accessible, which implies that all single qubit operations are possible. 
        This can be seen by inspecting the effective Hilbert spin space. 
        The isolated sub blocks of Eq.~(\ref{eqn:generalized_djjb_hamiltonian_subblock_a}) and Eq.~(\ref{eqn:generalized_djjb_hamiltonian_subblock_b}) belong to the SO(2) group. 
        When substituted into the unitary propagator of Eq.~(\ref{eqn:unitary-time-evolution-operator-hamiltonian-time-independent}), the resulting operation is SU(2). 
        The adjoint representation of SU(2) is isomorphic to SO(3). 
        Thus if one is able to realize all SO(3) operations on the Bloch vector, as has been shown, all relevant SU(2) operations can be realized.

    \subsection{Application to Ultra-Cold Optical Lattices}
        Because of the general nature of the total Hamiltonian considered in this paper, there are several immediate applications within the condensed matter context. 
        For example, Ref.~\cite{chung2021} recently found experimentally-realized resonance conditions of magnetic anisotropy and exchange coupling for a Mott insulator composed of an ultra-cold optical lattice of \ce{^{87}Rb} atoms.
        The authors uncover an oscillation in an observable parameter $A$, which represents the longitudinal spin alignment, that is maximized for the resonance condition of $J=\pm\tfrac{2}{3}D$. 
        Recovering the authors' form of the two-site Bose-Hubbard Hamiltonian used to explain these resonances from our model is trivial, and is accomplished by setting $J_{K2}=J_{K3}=0$, while noting the difference in the sign convention of $J$. 
        We use the experimental example to uncover another possible resonance, but also highlight a constraint within DJ resonances.
        Using the Hamiltonian $m=0$ sub-block in Ref.~\cite{chung2021}, if one were to prepare the system as a pure mixture of either state within the block, i.e. either state in $\{(\ket{1,-1}+\ket{-1,1})/\sqrt{2},\ket{0,0}\}$, one finds a resonance condition of $J=2D$.
        If one were to choose instead the product basis, a single sub block of the Hamiltonian would emerge associated with the $\{\ket{2,0},\ket{0,0}\}$ product states. 
        Preparation as a pure mixture of one of these states results in a different DJ resonance condition, namely $J=\frac{2}{9}D$.
        Whether any particular resonance is important experimentally, however, is the constraint on whether a measurement of the oscillations between states is feasible.
        
        It is interesting to note that applying our three-particle model with similar conditions leads to strikingly different behavior than the two-site Bose-Hubbard model. 
        To do this, one should first modify our model by matching the particle type across all three particles, i.e. change $S_{1}$ so that $S_{1}=S_{2}=S_{3} \equiv S_{1,2,3}=1$, and extend magnetic anisotropy to particle 1. 
        In doing so, and assuming an isotropic coupling between nearest neighbors (i.e. $J_{K2}=J_{H}=J$ and $J_{K3}=0$), the effective Hamiltonian becomes,
        \begin{align}
        \label{eqn:bose-hubbard-three-site}
            \mathcal{H}_{\text{BH3}} = J\left(\hat{\vb{S}}_{1}\cdot\hat{\vb{S}}_{2}+\hat{\vb{S}}_{2}\cdot\hat{\vb{S}}_{3}\right)+D\sum_{i=1}^{3}\hat{S}^{z}_{i}\hat{S}^{z}_{i}.
        \end{align}
        Within this total Hamiltonian, there are three two-dimensional blocks that correspond with transitions between different state phases: $\{\ket{2,2}_{A},\ket{2,2}_{B}\}$, $\{\ket{2,0}_{A},\ket{2,0}_{B}\}$, and $\{\ket{2,-2}_{A},\ket{2,-2}_{B}\}$.
        These blocks share the same form,
        \begin{align}
        \label{eqn:bose-hubbard-three-site-block}
            \mathcal{H}=\frac{1}{2}\left(J\sigma_{z}+\sqrt{3}J\sigma_{x}\right).
        \end{align}
        Application of the Rabi formula on this block gives a Rabi frequency $\Omega = \abs{J}$, and a transition probability amplitude of,
        \begin{align}
        \label{eqn:bose-hubbard-three-site-transitions}
            P(t)=\frac{3}{4}\sin^{2}\left(\Omega t\right).
        \end{align}
        Thus in the extension of the approximate form of the two-site Bose-Hubbard Hamiltonian to three sites, there is no DJ resonance within that basis, and the transition probability maxima is $3/4$. In contrast, we have found that the DJ resonances appear when considering exchange interactions that extend either to next-nearest neighbor exchange interactions (in the linear chain geometry) or in trimer geometries for $S=1$.

    \subsection{Application to Magnetic Molecules and Quantum Dots}
        The results of this work can be applied to magnetic molecule models. 
        In single-molecule quantum dots, like those explored in Ref.~\cite{hiraoka17}, the source of magnetic anisotropy is from spin-orbit interactions and the geometry of the system resembles a two-terminal molecular transistor. 
        Even so, the spin model approximation of the system resembles the spin model considered here. 
        Competing effects such as additional exchange coupling interactions with the substrate and decay mechanisms such as spin lattice vibrations could mask these DJ resonances. 
        As found in Ref.~\cite{switzer2021}, however, significant anisotropic application of the exchange coupling between particle 1 and the others still allows for usable entangled state switching (up to P=0.995) around a DJ resonance. 
        This is important when considering the experimental uncertainties inherent in a scenario of coupling an electron to a magnetic dimer on a substrate.
        Furthermore, the existence of a DJ resonance for any value of $S_{2,3}$ via Eq.~(\ref{eqn:dj-resonance-general-s}) opens the application of this model to larger spin magnetic molecules coupled together, such as coupled $\text{Mn}_{3}$ monomers.

        When considering a trimer geometry for the $S_{1,2,3}=1$ model, or equivalently a next-nearest neighbor exchange interaction in a chain geometry, the preceding section's assumption of $J_{\text{K3}}=0$ is modified.
        Assuming an isotropic application of the exchange strength coupling between each particle in the three-particle system, the $S_{1,2,3}=1$ model's effective spin Hamiltonian is
        \begin{align}
            \nonumber
            \mathcal{H}_{\text{trimer}} &= J\left(\hat{\vb{S}}_{1}\cdot\hat{\vb{S}}_{2}+\hat{\vb{S}}_{2}\cdot\hat{\vb{S}}_{3}+\hat{\vb{S}}_{3}\cdot\hat{\vb{S}}_{1}\right)
            \label{eqn:trimer-hamiltonian}
            \\&\;\;\;+D\sum_{i=1}^{3}\hat{S}^{z}_{i}\hat{S}^{z}_{i}.
        \end{align}
        In the device basis, this Hamiltonian contains four two-dimensional blocks corresponding with the conserved spin transitions of $m=\pm1$ and $m=\pm2$. 
        The $m=\pm1$ blocks contain the interactions of the $\left\{\ket{1}\ket{1,0},\ket{0}\ket{1,1}\right\}$ and $\left\{\ket{0}\ket{1,-1},\ket{-1}\ket{1,0}\right\}$ states, respectively. 
        The $m=\pm1$ block's effective form is,
        \begin{align}
        \label{eqn:trimer-device-basis-m1}
            \mathcal{H}^{\text{trimer}}_{m=\pm1}=\pm D \sigma_{z}+J\sigma_{x}.
        \end{align}
        The $m=\pm2$ blocks involve the $\left\{\ket{1}\ket{2,1},\ket{0}\ket{2,2}\right\}$ and $\left\{\ket{0}\ket{2,-2},\ket{-1}\ket{2,-1}\right\}$ states, respectively, and have the form,
        \begin{align}
        \label{eqn:trimer-device-basis-m2}
            \mathcal{H}^{\text{trimer}}_{m=\pm2}=\pm \frac{1}{2}J \sigma_{z}+\sqrt{2}J\sigma_{x}.
        \end{align}
        Application of the Rabi formula to both the $m=\pm1$ and $m=\pm2$ blocks yields two cases. 
        When the state transition is in the $m=\pm1$ block, the Rabi frequency is $\Omega_{\pm1}=\sqrt{D^{2}+J^{2}}$ with a transition probability amplitude of,
        \begin{align}
        \label{eqn:trimer-device-basis-transitions-m1}
            P_{\pm1}(t)=\left(\frac{J}{\Omega_{\pm1}}\right)^{2}\sin^{2}(\Omega_{\pm1} t).
        \end{align}
        In other words, the two state system possesses a DJ resonance at $D=\pm J$.
        If instead the state transition is in the $m=\pm2$ block, the Rabi frequency is $\Omega_{\pm2}=\tfrac{3}{2}\abs{J}$ with an amplitude of,
        \begin{align}
        \label{eqn:trimer-device-basis-transitions-m2}
            P_{\pm2}(t)=\frac{8}{9}\sin^{2}(\Omega_{\pm2} t).
        \end{align}
        Thus in the $m=\pm2$ block, there is no DJ resonance, and the maximum transition probability amplitude is $8/9$. 
        This maximum matches the maximum amplitude in the $S_{1,2,3}=\tfrac{1}{2}$ model where anisotropy does not play a role.
        We also note that in the trimer model, another DJ resonance exists within a different basis: the product basis. 
        In that basis, there are two two-dimensional blocks that contain the transitions between the $\ket{2,\pm1}_{B}$ and $\ket{1,\pm1}_{B}$ states. 
        They take the form,
        \begin{align}
        \label{eqn:trimer-hamiltonian-product-basis-m1}
            \mathcal{H}^{\text{trimer}}_{\pm1}=J\sigma_{z}\pm D\sigma_{x}.
        \end{align}
        We note the similarity to Eq.~(\ref{eqn:trimer-device-basis-m1}), but with the reversed role of $D$ and $J$ on Bloch vectors within the corresponding Bloch sphere. 
        The Rabi frequency is the same ($\Omega_{\pm1}=\sqrt{D^2+J^2}$), and the transition probability amplitude reflects the $D$ and $J$ role reversal,
        \begin{align}
        \label{eqn:trimer-product-basis-transitions-m1}
            P_{\pm1}(t)=\left(\frac{D}{\Omega_{\pm1}}\right)^{2}\sin^{2}(\Omega_{\pm1} t).
        \end{align}
        
    \subsection{Summary}
        In summary, we have demonstrated the general existence formula for DJ resonances and how these resonances enable full control of the Bloch vector in an appropriately chosen Bloch sphere representation. 
        We also contrasted the difference in entanglement switching mechanisms for the spin $S_{2,3}=\tfrac{1}{2}$ and $S_{2,3}=1$ models. 
        We have shown that in order to achieve acceptable control of a Bloch vector within the $S_{2,3}=\tfrac{1}{2}$ model, additional mechanisms such as outgoing spin filters for particle 1 or anisotropy of the exchange interaction between particle 2 and 3 is required. 
        Last, we have indicated several systems of interest where DJ resonances could be used to explore qubit scenarios and to discover fundamental phenomena.

\begin{acknowledgments}
    We thank James Freericks, Peter Dowben, Volodymyr Turkowski, Silas Hoffman, and Dave Austin for helpful discussions. This work was supported by the Center for Molecular Magnetic Quantum Materials, an Energy Frontier Research Center funded by the U.S. Department of Energy, Office of Science, Basic Energy Sciences under Award No. DE-SC0019330. The authors declare no competing financial interests.
\end{acknowledgments}

\bibliography{biblio}
\end{document}